\documentclass[aps,prl,amsmath,amssymb,twocolumn]{revtex4}
\pdfoutput=1
\usepackage{graphicx}
\usepackage{color}
\newcommand{\beq}{\begin{eqnarray}}
\newcommand{\eeq}{\end{eqnarray}}

\newcommand{\bmp}{\noindent\begin{minipage}{16cm}}
\newcommand{\emp}{\end{minipage}\vskip 7mm} 

\usepackage{dcolumn}
\usepackage{bm}
\usepackage{bbm}
\usepackage{subfigure}
\usepackage{pxfonts}

\usepackage{epsfig}

\def\lsim{\mathrel{\rlap{\lower4pt\hbox{\hskip1pt$\sim$}}
    \raise1pt\hbox{$<$}}}                
\def\gsim{\mathrel{\rlap{\lower4pt\hbox{\hskip1pt$\sim$}}
    \raise1pt\hbox{$>$}}}                

\newcommand{\be}{\begin{eqnarray}}
\newcommand{\ee}{\end{eqnarray}}

\begin{document}
\title{\Large  \color{red} Mass Deformed Exact S-parameter in Conformal Theories}
\author{Francesco {\sc Sannino}$^{\color{blue}{\varheartsuit}}$}
\email{sannino@cp3.sdu.dk} 
\affiliation{
$^{\color{blue}{\varheartsuit}}${ CP}$^{ \bf 3}${-Origins}, 
Campusvej 55, DK-5230 Odense M, Denmark.\footnote{CP$^3$- Origins: 2010-21}}
\begin{abstract}
We  use the exact expression for the S parameter in the perturbative region of the conformal window to establish its dependence on the explicit introduction of fermion masses. We demonstrate that the relative ordering with which one sends to zero either the fermion mass or the external momentum leads to drastically different limiting values of S. Our results apply to any fermion matter representation and can be used as benchmark for the determination of certain relevant properties of the conformal window of any generic vector like gauge theory with fermionic matter. We finally suggest the existence of a universal lower bound on the opportunely normalized S parameter and explore its theoretical and phenomenological implications. Our precise results constitute an ideal framework to correctly interpret the lattice studies of the conformal window of strongly interacting theories. \end{abstract}

\maketitle

Depending on the number of flavors, matter representation and colors, 
non-abelian gauge theories are expected to exist in a 
number of distinct phases, classifiable according 
to the force felt between two static sources.  
The knowledge of this phase diagram is relevant 
for the construction of extensions of the 
Standard Model (SM) that invoke dynamical electroweak symmetry
breaking \cite{Weinberg:1979bn,Susskind:1978ms}.   An up-to-date review is \cite{Sannino:2009za} while earlier reviews are \cite{Hill:2002ap,Yamawaki:1996vr}.  The phase diagram is also useful in providing ultraviolet completions of 
unparticle \cite{Georgi:2007ek} models \cite{Sannino:2008nv,Sannino:2008ha} and it has been investigated recently using different analytical methods  \cite{Sannino:2004qp}. 

The goal here is to determine, in an exactly calculable regime, the effects on the precision parameters due to the introduction of an explicit mass term for fermions in a gauge theory featuring a true IRFP. Our results are important for the explorations of the conformal window since they shed light on relevant properties in this region. The results are also directly applicable to unparticle extensions of the SM \cite{Georgi:2007ek,Sannino:2008nv}. 

 We stress, to avoid any possible misunderstanding, that the goal of this paper is to investigate the effects on the vacuum polarizations associated to a {\it perturbative} conformal sector in presence of a mass deformation. The language of the electroweak precision parameters is borrowed to connect more easily to the phenomenological world. In the first part of this work, which concerns exact results, we will not address the breaking of the electroweak symmetry and hence we choose the reference of the Higgs mass in such a way that the sole contributions to the precision parameters come from the calculable new sector. In other words we are computing the $VV-AA$ two-point function in isolation. However, in the last part of the paper we propose how to make contact with earlier model computations in the near conformal regime.

The relevant corrections due to the presence of new physics trying to modify the electroweak breaking sector of the SM appear in the vacuum polarizations of the electroweak gauge bosons. These can be parameterized in terms of the three 
quantities $S$, $T$, and $U$ (the oblique parameters) 
\cite{Peskin:1990zt,Peskin:1991sw,Kennedy:1990ib,Altarelli:1990zd}, and confronted with the electroweak precision data. We will concentrate on the first two. 

The oblique parameters $S$ and $T$ definitions we use here are the one as in \cite{He:2001tp}:
 \be S&=&-16\pi\frac{\Pi_{3Y}(m_Z^2)-\Pi_{3Y}(0)}{m_Z^2} \,,  \\
  T&=&4\pi\frac{\Pi_{11}(0)-\Pi_{33}(0)}{s_W^2c_W^2m_Z^2} \,, 
\ee
where the weak-mixing angle $\theta_W$ is defined at the scale $\mu =m_Z$. 
$\Pi_{11}$ and  $\Pi_{33}$ are the vacuum polarizations of isospin currents, and $\Pi_{3Y}$ the vacuum polarization of one isospin and one hypercharge current. Here we use as reference point, instead of the $Z_0$ mass, the external momentum $q^2$. The value assumed by these parameters depend on the specific extension of the SM.

The fundamental differences with respect to earlier investigations \cite{Appelquist:1998xf,Kurachi:2006mu} are that: 
\begin{itemize}
\item Our results are under perturbative control since we consider a sufficiently large number of flavors so that the IRFP occurs at a  perturbative value of the underlying gauge theory coupling constant. In fact we are using the non interaction limit. It would be interesting to extend the computation to the next leading order in the new gauge theory coupling constant. 

\item We consider two independent limits in momentum space, the one in which the external momentum vanishes keeping fixed the fermion masses, and the opposite limit in which the mass term vanishes at a finite value of the external momentum. 
  \end{itemize}
 
Lattice simulations of the properties of the conformal window \cite{Catterall:2007yx}  can now test their results against the predictions made here in a controllable way.  
 
 We start by considering a gauge theory with sufficient fermionic matter to develop a perturbative IRFP. Let's define with $N_f$ the number of Dirac flavors of this theory.  This is the Banks-Zaks \cite{Banks:1981nn} regime of the gauge theory. The associated quantum global symmetries are 
 $SU_L(N_f)\times SU_R(N_f)\times U_V(1)$ if the fermion representation is complex or $SU(2N_f)$ if real or pseudoreal. To make contact with the SM we imagine to weakly gauge $N_D=N_f/2$ doublets. If any gauge anomaly arises, with respect to the SM interactions, one can always add new fermion doublets, neutral with respect to the new dynamics to cancel the anomalies. 
  
We give to the up and down type fermions, with respect to the electroweak interactions, masses $M_1$ and $M_2$. All the up and down fermions have the same mass.  To leading order in the new gauge coupling, in the case of Dirac-type masses for the fermions, and at the one-loop level the expressions for the oblique parameters read
\cite{He:2001tp}:
\begin{eqnarray}
S&= & \frac{\sharp}{6\pi}
\left\{
2(4Y+3)x_1+2(-4Y+3)x_2-2Y\ln\frac{x_1}{x_2}
\right.\nonumber \\[1mm]
&&\left. +\left[\left(\frac{3}{2}+2Y\right)x_1+Y\right] G(x_1)  +\left[\left(\frac{3}{2}-2Y\right)x_2-Y\right] G(x_2) \right\}\,, \label{eq:Sfermion}\nonumber \\[1mm] &&\\[1mm]
T&=&\frac{\sharp}{8\pi{s}_W^2c_W^2}F(x_1,x_2) \,, \label{eq:Tfermion}
 \end{eqnarray} where  $x_i=(M_i/m_Z)^2$ 
with $i=1,2$ and  $\sharp = N_D \, d[r] $ counts the number of doublets times the dimension of the representation ($d[r]$) under which the fermions transform. {}For example for the fundamental representation $d[F] = N$, for an $SU(N)$ gauge group and $d[S] = N(N+1)/2$ for the 
two-index symmetric representation of 
the gauge group. 

The functions $G(x)$ and  $F(x_1,x_2)$ are defined via:
\begin{eqnarray}
F(x_1,x_2)&=&\frac{x_1+x_2}{2}-\frac{x_1x_2}{x_1-x_2}\ln\frac{x_1}{x_2}
\,,
\label{eq:Ffun}
\\[2mm]
G(x)&=&-4\sqrt{4x-1}\,\arctan\frac{1}{\sqrt{4x-1}}
\,,
\label{eq:Gfun}
\end{eqnarray}
We replace $m_Z^2$ with the momentum $q^2$ and consider different limits with respect to the masses of the new fermions belonging to a perturbative conformal field theory at large distances. 
We note that at the one loop level, presented here for the vacuum polarizations, the new gauge dynamics, assumed to be perturbative, does not contribute.

\vskip .2cm
\noindent
{\bf Sending $q^2$ to zero keeping fixed the fermion masses:}
\vskip .2cm
\noindent
We assume $M_1=M_2 = m$ and obtain: 
\begin{eqnarray}
\lim_{\frac{q^2}{m^2}\rightarrow 0}S =  \frac{\sharp}{6\pi}\left[1 + \frac{1}{10x} + \frac{1}{70 x^2} + {\cal O}(x^{-3})\right] \ , 
\label{smallq}
\end{eqnarray}
with $ x=\frac{m^2}{q^2}$. 
Note that for $M_1=M_2=m$ the dependence on the hypercharge $Y$ vanishes. We shall be concerned with this limit here but we included the general formulae since they will be useful when considering non-degenerate fermions or when giving different masses only to a certain number of doublets. We observe that, in perturbation theory, when the new gauge interactions are weak, the leading term at zero momentum does not depend on the explicit value of the fermion masses. Turning on a momentum smaller than $m$ we note that the leading dependence is proportional to the inverse of the mass parameter squared. 

The $T$ and $U$ parameters are zero by weak isospin invariance in the $x_1=x_2$ limit which corresponds to the case $M_1=M_2=m$. 

It is important to observe that in the $q^2/m^2 \rightarrow 0$ limit the $S$ parameter does {\it not} vanish but it becomes a pure number!  At first, this seems a surprising result given that one often hears the {\it naive} statement: {\it The $S$ parameter vanishes in the conformal region}. 
\begin{figure}[htbp]
\begin{center}
\includegraphics[width=.45\textwidth]{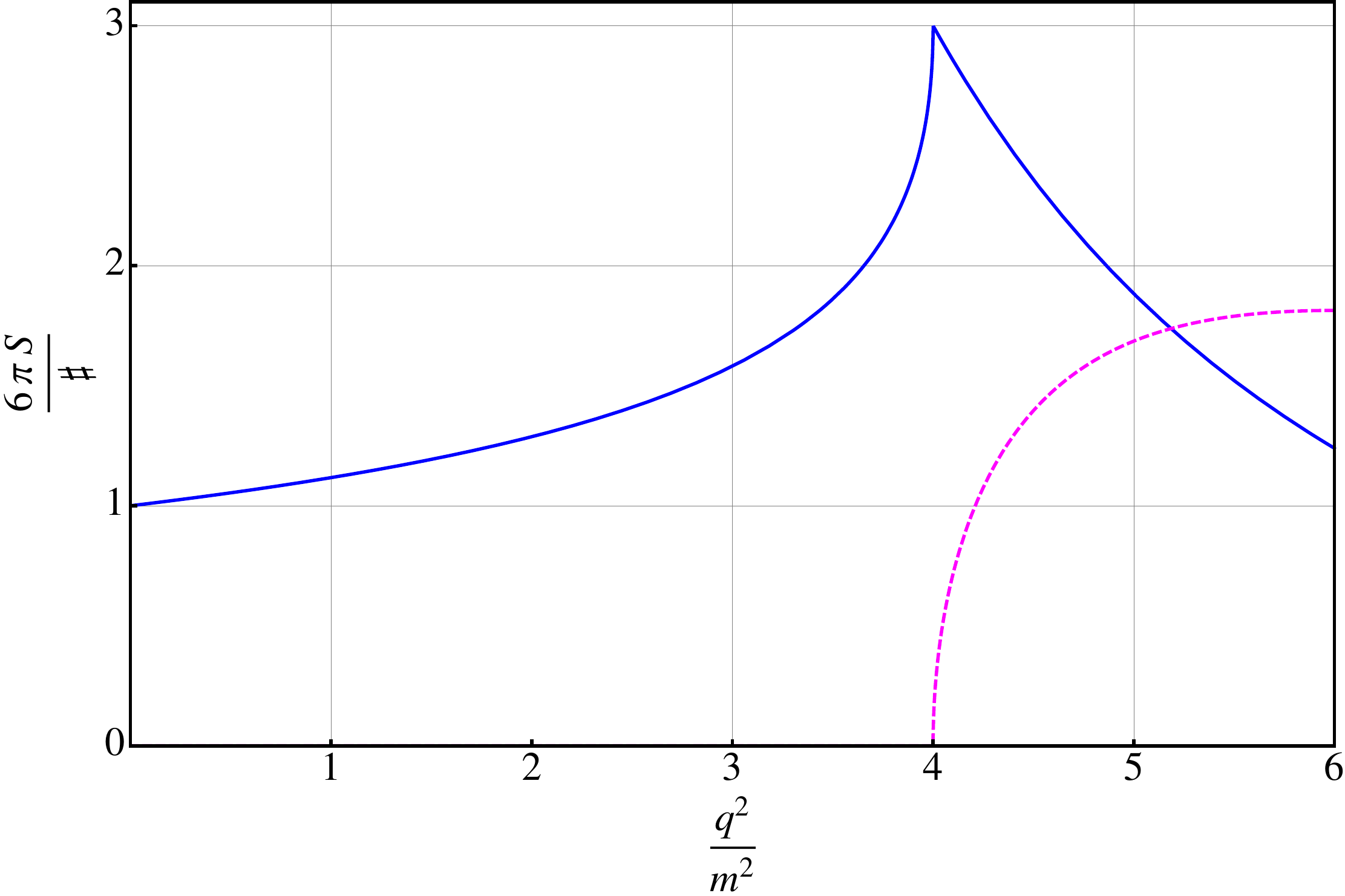} 
\caption{Real (blue-solid) and imaginary (magenta-dashed) parts for the normalized $\displaystyle{\frac{6\pi S}{\sharp}}$ parameter as function of increasing $q^2/m^2$ and $\sharp = \frac{N_f}{2}\, d[r] $.} 
\label{Sqlimit}
\end{center}
\end{figure}%
The point is that we have not yet approached the conformal limit. This is achieved only when we first send to zero the fermion mass while keeping the momentum finite. 
In Fig.~\ref{Sqlimit} we plot the perturbatively exact real (blue-solid) and imaginary (magenta-dashed) parts of the opportunely normalized $S$ parameter, i.e. $\displaystyle{\frac{6\pi S}{\sharp}}$, near the perturbative fixed point, as function of increasing $q^2/m^2$. We observe that for small momenta the normalized $S$ parameter decreases to unity while at $q^2 = 4m^2$ an imaginary part develops and steeply rises while the real part starts decreasing. This is the kinematical cut associated to particle production in the fermion loop since the external momentum is sufficiently large to create, on shell, a fermion-antifermion pair. Clearly the Taylor-expanded formula in \eqref{smallq} applies below the production threshold and for very small $q^2/m^2$.

\vskip .2cm
\noindent
{\bf Sending $m^2$ to zero first. The conformal limit:}
\vskip .2cm
\noindent
Expanding around this limit we find: 
\begin{eqnarray}
\lim_{\frac{m^2}{q^2}\rightarrow 0} \Re[S] & = &  x\,\frac{\sharp}{\pi}\left[2 + \log (x) \right] + {\cal O}(x^{2}) \ , \label{real}  \\
\lim_{\frac{m^2}{q^2}\rightarrow 0} \Im[S] & = &  x\,\sharp + {\cal O}(x^{2})   \  .
\label{ima}
\end{eqnarray}
The imaginary and real parts of the $S$ parameter are both nonzero and vanish with the mass when keeping fixed the external reference momentum $q^2$. 
\begin{figure}[htbp]
\begin{center}
\includegraphics[width=.45\textwidth]{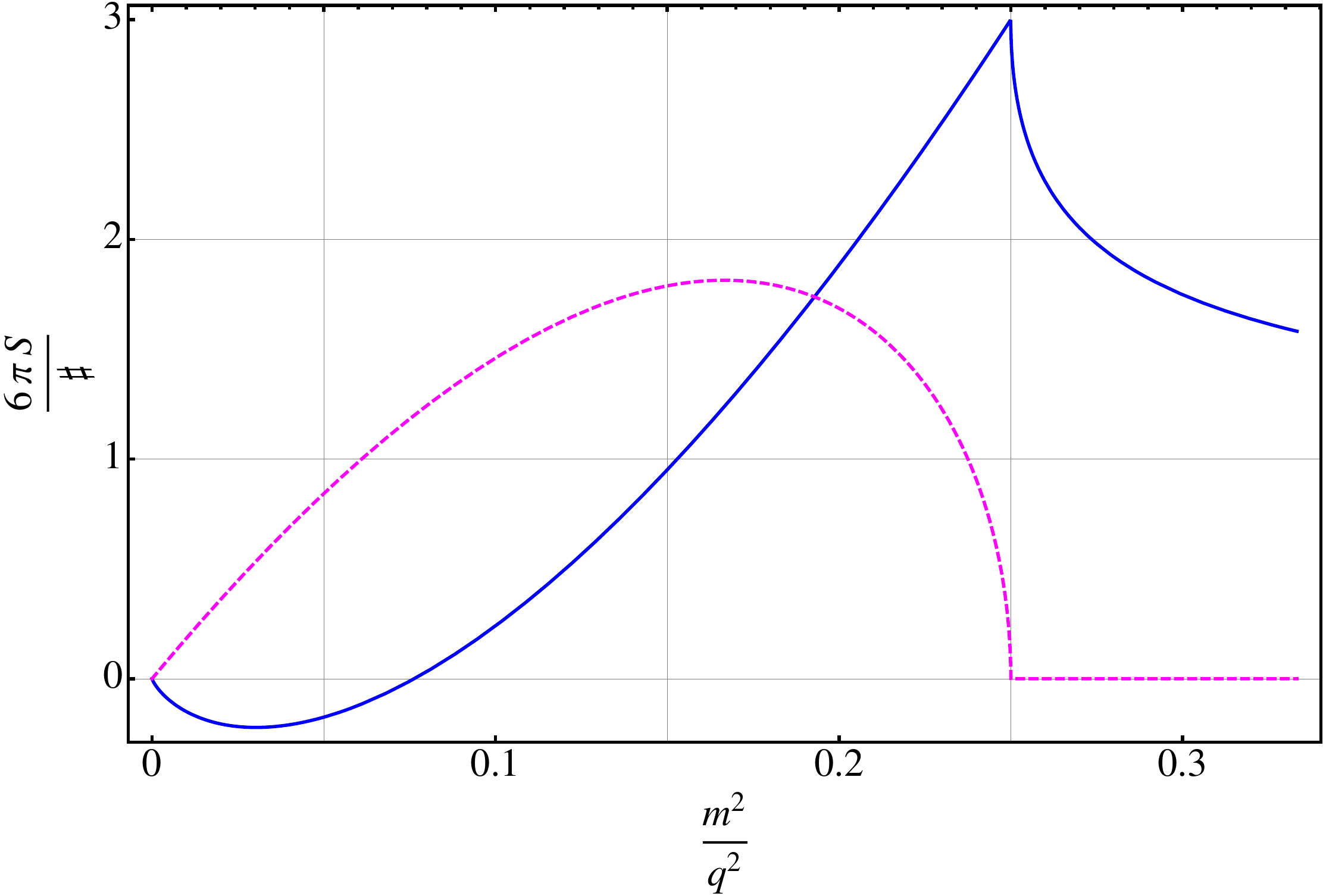}
\caption{Real (blue-solid) and imaginary (magenta-dashed) parts for the normalized $\displaystyle{\frac{6\pi S}{\sharp}}$ parameter as function of increasing $m^2/q^2$ and $\sharp = \frac{N_f}{2}\, d[r] $ . }
\label{Smlimit}
\end{center}
\end{figure}%
We plot in Fig.~\ref{Smlimit} the real (blue-solid) and imaginary (magenta-dashed) parts  for the normalized $\displaystyle{\frac{6\pi S}{\sharp}}$ parameter as function of increasing $m^2/q^2$. We note that due to the logarithmic term the $\Re[S]$ becomes {\it negative} before approaching zero. The imaginary part rises linearly, at first, as shown in \eqref{ima} but then goes to zero at $4 m^2 =q^2$ as dictated by kinematics. We are finally approaching the conformal limit. 

In the perturbative conformal regime the fermion masses leading contribution to the condensate $\langle \bar{\psi} \psi \rangle$ is linear, i.e. $\langle \bar{\psi} \psi \rangle \propto m \Lambda_U^2$ \cite{Sannino:2008nv,Sannino:2008pz} and $\Lambda_U$ os a conveniently defined ultraviolet scale. There can be subleading  non-analytic terms in the mass due to the effects of the axial anomaly \cite{Sannino:2008pz} or due to the occurrence of critical exponents \cite{Sannino:2008nv,DelDebbio:2010ze}. We can always trade, in the formulae above, the fermion mass for the fermion condensate.

\vskip .2cm
\noindent
{\bf S-parameter lower bound:}
\vskip .2cm
\noindent
What happens when we decrease the number of flavors within the conformal window? The theory becomes more and more non-perturbative and our results receive relevant corrections.  As we further decrease the number of flavors we cross into the chirally broken phase and conformality is lost. Below the critical number of flavors a dynamical mass of the fermions generates. In the broken phase we should compute the $S$ parameter, in the zero momentum limit, with the hard mass of the fermions replaced by the hard plus the dynamical one. We can use as straightforward estimate for $S$ the {\it naive} one loop approximation. We clearly recover, for the normalized $S$ at zero momentum, the unity value. This suggests that the broken and symmetric phases are smoothly connected.

A further natural expectation,  is that the normalized $S$ (in the same kinematical limit), is non-decreasing, with respect to the unity value, as we decrease the number of flavors across the entire phase diagram. This expectation is supported by the fact that, for Quantum Chromo Dynamics (QCD), in the chirally broken phase the normalized physical $S$ is about 1.9 times the unity value \cite{Peskin:1991sw}. Of course, since we are assuming now that chiral symmetry corresponds to the breaking of the electroweak symmetry, we also need to set the reference value of the Higgs mass \cite{Peskin:1991sw}.  We recall that in the literature for beyond standard model physics one typically assumes the reference momentum to be smaller than the physical scale associated to the new physics  when approaching the conformal window from the broken phase. 

We thereby suggest that the $S$ parameter satisfies the following bound:
 \begin{equation}
S_{\rm norm} \equiv \frac{6\pi S}{\sharp} \geq 1  \quad {\rm when} \quad {\frac{q^2}{m^2} \rightarrow 0}  \ ,
\label{sbound}
\end{equation}
when we send to zero the external momentum at a nonzero value of the fermion mass, either explicit or dynamical. We conjecture this to be the lower bound on the normalized $S$ parameter in the conformal window and in the chirally broken phase.  

The presence of a lower bound does not contradict  the statements made earlier in the literature that the $S$ parameter in near conformal theories can be smaller than the one in QCD. A reduction of $S_{\rm norm}$ parameter with respect to the QCD value is possible but should not violate the bound \eqref{sbound}.

To make contact with earlier model computations we take the $S$ parameter estimate in near conformal theories deduced in \cite{Appelquist:1998xf} and generalized for arbitrary representations in \cite{Foadi:2007ue}: 
\begin{equation}
S = 4\pi \, F^2_{\pi} \left[ \frac{1}{M^2_V} +\frac{1}{M^2_A} - a\,\frac{8\pi^2}{d[r]}  
\frac{F^2_{\pi}}{M^2_V \, M^2_A} \right] \ .
\end{equation}
$F_{\pi}$ is the pion decaying constant, $M_{V/A}$ are the masses of the lightest vector/axial spin one states, and the last term encodes the effects of the heavier but closely spaced states which cannot be neglected in a near conformal theory. The unknown nonuniversal parameter $a$ is expected to be greater or equal to zero. This expression is derived exactly in the $q^2\rightarrow 0$ limit and hence we can now confront it with our bound. It is clear that, for $M \approx 2\pi F_{\pi}/d[r]$, the value of $a$ can be near unity without violating the bound. 

The lower bound has a direct impact on the construction of models of dynamical electroweak symmetry breaking since it shows that models with the smallest number of techniflavors  gauged under the electroweak symmetry are favored by precision tests \cite{Sannino:2004qp, Dietrich:2005jn,Foadi:2007ue}. 

If we remain in the conformal window, while decreasing the number of flavors, but considering the limit of small masses and fixed external momentum, straightforward scaling relations tell us that $S \propto 4\pi F^2_{\pi} /q^2$.  This expression is consistent with the result in \eqref{real}, up to logarithmic corrections. This latter expression, however, generalizes also to the non-perturbative regime and shows that the $S$ parameter, in this kinematical limit, scales to zero with the same critical exponents of $F_\pi^2$ as function of the fermion mass.  The limit of sending to zero the mass first is important for models of unparticle physics.

\vskip .5cm
 The exact results presented here provide a natural benchmark for lattice computations of the $S$ parameter for vector-like gauge theories featuring an IRFP.  We have shown that if the external momentum vanishes, at a finite value of the fermion mass, the normalized $S$ parameter is real and  {\it decreases} to unity.  However, in the opposite regime, i.e. when the mass of the fermions tends to zero while keeping fixed the external momentum, the real and imaginary parts of the $S$ parameter vanish as well. Interestingly before reaching the zero value we discover that the real part is {\it negative}. This feature is interesting for models of unparticle physics.  We also proposed the existence of a universal lower bound for the normalized $S$ parameter in a given kinematical regime and discussed theoretical and phenomenological implications.

\acknowledgments
It is a pleasure  to thank Luigi Del Debbio for pleasant discussions at 3500 mt  elevation on the Aspen Mountains, Robert Shrock and  Rohana Wijewardhana for discussions, valuable comments and careful reading of the manuscript. We also thank the Aspen Center for Physics for the hospitality and its support while this work was completed.

\end{document}